\documentclass[twocolumn,pra,aps,showpacs,superscriptaddress]{revtex4}

\usepackage{amssymb}

\usepackage[english]{babel}
\usepackage{latexsym}
\usepackage{graphics}
\usepackage{subfigure}
\usepackage{epsfig}
\usepackage{graphicx}
\usepackage{dcolumn}
\usepackage{amsmath}
\usepackage{bm}

\newcommand{\be}{\begin{equation}}
\newcommand{\ee}{\end{equation}}
\newcommand{\bey}{\begin{eqnarray}}
\newcommand{\eey}{\end{eqnarray}}
\newcommand{\bw}{\begin{widetext}}
\newcommand{\ew}{\end{widetext}}

\begin{document}

\title{Self-trapping of Bose-Einstein condensates in optical lattices}
\author{Bingbing Wang}
\affiliation{Institute of Physics, 
Chinese Academy of Sciences, Beijing 100080, China}
\author{Panming Fu}
\affiliation{Institute of Physics, 
Chinese Academy of Sciences, Beijing 100080, China}
\author{Jie Liu}
\affiliation{Institute of Applied Physics and Computational
Mathematics, Beijing 100088, China}
\author{Biao Wu}
\email{bwu@aphy.iphy.ac.cn}
\affiliation{Institute of Physics, Chinese Academy of Sciences,
Beijing 100080, China}

\begin{abstract}
The self-trapping phenomenon of Bose-Einstein condensates (BECs) in
optical lattices is studied extensively by numerically solving the 
Gross-Pitaevskii equation. Our numerical results not only
reproduce the phenomenon that was observed in a recent experiment
[Anker {\it et al.}, Phys. Rev. Lett. {\bf 94} (2005)020403], but
also find that the self-trapping breaks down at long evolution 
times, that is, the self-trapping in optical lattices
is only temporary. The analysis of our numerical results shows
that the self-trapping in optical lattices is related to
the self-trapping of BECs in a double-well potential. A possible
mechanism of the formation of steep edges in the wave packet evolution
is explored in terms of the dynamics of relative phases between
neighboring wells.
\end{abstract}


\maketitle

\section{Introduction}
Progress in recent years has shown that a Bose-Einstein condensate(BEC)
in an optical lattice is a fascinating periodic system, where the physics
can be as rich as in fermionic periodic systems, the main subject of
condensed-matter physics. In such a bosonic system, people have observed 
well-known and long predicted phenomena, such as Bloch oscillations
\cite{Morsch2001PRL} and the quantum phase transition between 
superfluid and Mott-insulator\cite{Greiner2002Nature1}.
More importantly, there are new phenomena that have been
either observed or predicted in this system, for example,
nonlinear Landau-Zener tunneling between Bloch bands
\cite{NlzExp2003PRL,WuAndNiu2000} and the strongly inhibited
transport of one dimensional BEC in an optical 
lattice\cite{Fertig2005PRL}.

\begin{figure}[!tb]
\includegraphics[width=\columnwidth]{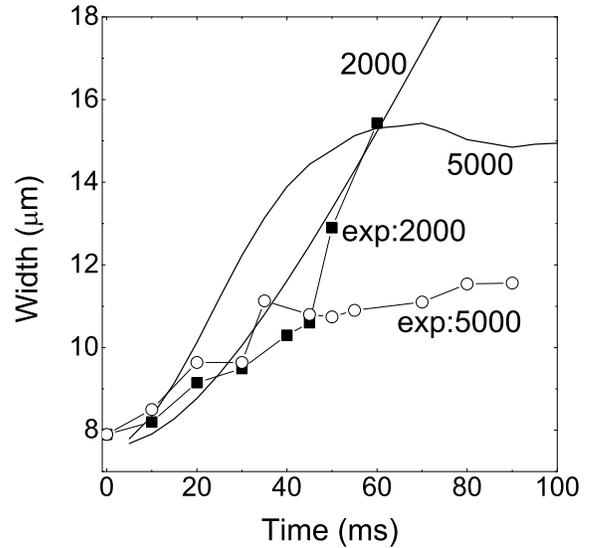}
\caption{The width of the BEC wave packet as a function of
time for $N=2000$ and $N=5000$. $N$ is the number of atoms
in the BEC. The solid lines are our numerical results while
the circles and squares are experimental data 
from Ref.\cite{Anker2005PRL}.}
\label{fig:width}
\end{figure}

Another intriguing phenomenon, self-trapping, was recently observed
experimentally in this system\cite{Anker2005PRL}. In this experiment,
a BEC with repulsive interaction was first prepared in a dipole trap. 
By adiabatically ramping up an optical lattice, the BEC was essentially 
transformed into a Bloch state at the center of the Brillouin zone. 
With the optical lattice always on, the BEC was then released into 
a trap that serves as a one dimensional waveguide. The evolution of 
the BEC cloud inside the combined potential was studied by taking 
absorption images. When the number of atoms in the BEC is small, 
say around 2000, the BEC wave packet
was found to expand continuously (which is expected). However,
when the number of atoms was increased to about 5000, it was observed
that the BEC cloud stops to expand after initially expanding
for about 35ms (see Fig.\ref{fig:width}). This is quite 
counter-intuitive. Without interaction, a wave packet with a narrow 
distribution in the Brillouin zone expands continuously inside 
a periodic potential. One would certainly expect that with a repulsive 
interaction between atoms the BEC cloud spread faster.  
This experiment showed the contrary: if the cloud is dense enough, 
it self-traps and stops spreading.

To understand this intriguing phenomenon, we have carried out
extensive numerical study of this system with the one-dimensional(1D)
Gross-Pitaevskii equation. Our results match quite well with the experimental
data as shown in Fig.\ref{fig:width}. When the atom number $N$
in the BEC is 2000, the agreement between our numerical results
and the experiment is excellent; when $N=5000$, our results
are about 40\% larger than the experimental data.
The discrepancy in the latter case is likely caused by the higher
density: with higher density the lateral motion of the BEC cloud
may become more relevant to the longitudinal expansion; 
however, the lateral motion is completely ignored in our numerical study 
as we use the 1D Gross-Pitaevskii equation. 
\begin{figure}[!htb]
\includegraphics[width=\columnwidth]{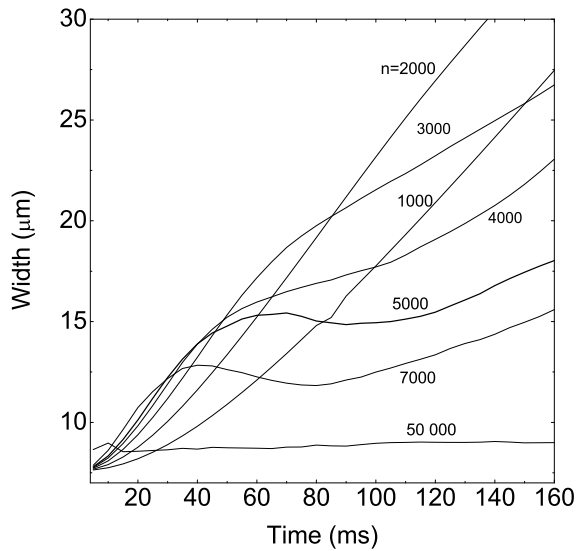}
\caption{The width of the BEC wave packet as a function of
time for $N=1000,2000,3000,4000,5000,7000$, and $50000$.}
\label{fig:width2}
\end{figure}

Very interestingly, we find that the self-trapping is temporary. 
After a sufficiently long evolution time, the self-trapping 
breaks down and the wave packet starts to expand again as seen in 
Fig.\ref{fig:width2}. Since the break-down time is much longer
than the observation used in the current experiment\cite{Anker2005PRL},
these results need to be verified in future experiments. 
This breakdown of self-trapping is likely caused by the leakage
of atoms at the outmost wells, or as we shall call it the dripping effect.
Furthermore, our numerical results show that the steep edges 
do not necessarily lead to the self-trapping and always appear 
in the wave packet evolution, independent of its denseness of 
the BEC cloud. This is different from Ref.\cite{Anker2005PRL}, where it 
was pointed out that the steep edges appearing at the two sides of 
the wave packet are crucial for the appearance of the self-trapping. 
The wave packet evolution in the quasi-momentum space is also studied.
We find that the wave packet localizes largely near the center of the
Brillouin zone; no major interesting features can
be identified during the evolution.

Besides, we have analyzed our numerical results in detail, in 
particular, in terms of the relative atom number difference and 
the relative phase difference between neighboring wells. With such an analysis,
we have confirmed the previous study\cite{Anker2005PRL} that 
the self-trapping observed in optical lattices is closely 
related to the self-trapping of a BEC in the double-well potential
\cite{Milburn1997PRA,Smerzi1997PRL,Albiez2005PRL}. From this
analysis, we have also explored a possible mechanism of the formation 
of steep edges.

Our paper is organized as follows. We shall first present a
brief description of our numerical method. We then describe
how the wave packets evolve in our numerical simulation. 
Afterwards, we analyze our numerical results in an attempt
to understand our numerical results. The analysis is
done from the angle of a BEC in a double-well potential.
Summary and some discussion are given at the end.

\section{Numerical method}
To model the experiment, we use the following Gross-Pitaevskii(GP)
equation
\begin{eqnarray}
&&i\hbar{\partial\over \partial t}\psi(\bm{r},t)=
-{\hbar^2 \over 2m}\nabla^2\psi(\bm{r},t)+
V_0\cos(2k_Lx)\psi(\bm{r},t)+\nonumber\\
&&~~~~~~+V_{wg}(\bm{r})\psi(\bm{r},t)+
\frac{4\pi\hbar^2a_s}{m}|\psi(\bm{r},t)|^2
\psi(\bm{r},t)\,,
\label{eq:gpe3}
\end{eqnarray}
where $m$ is the atomic mass, $a_s$ is the $s$-wave scattering
length, $k_L$ is the wavelength of the laser 
that generates the optical lattice, and $V_{wg}(\bm{r})$ describes 
the waveguide potential. Due to the tight confinement perpendicular to
the optical lattice from the waveguide potential,
the dynamics of this system is largely one-dimensional.
This allows us to integrate out the two perpendicular
directions and reduce the above GP equation to
\begin{eqnarray}
i{\partial \psi(x,t) \over \partial t}&=&-{1 \over 2}
\nabla^2\psi(x,t)+V\cos(x)\psi(x,t)+\nonumber\\
&+&{1 \over 2}\omega x^2\psi(x,t)+g|\psi(x,t)|^2\psi(x,t)\,,
\label{eq:gpe1}
\end{eqnarray}
were we have made the equation dimensionless. In doing
so, we have $x$ in units of $1/2k_L$ and $t$ in
$m/4\hbar k_L^2$. The strength of the optical lattice 
is given $V=V_0/16E_r$ with $E_r=\hbar^2 k_L^2/2m$ being the recoil
energy. For the nonlinear interaction, we have
\be
g={\pi a_s m \omega_\bot N
\over \surd 2 \pi \hbar k_L}\,,
\label{eq:nonlinear}
\ee
where $N$ the total number of the BEC in the harmonic trap
and $\omega_\bot$ is the transverse trapping
frequency of the waveguide. The other frequency $\omega$ is so
chosen that the initial rms-width of BEC wave packet is $7.6\mu$m 
as in the experiments\cite{Anker2005PRL}. This width
corresponds to about 100 wells occupied. The wave function
$\psi$ is normalized to one. In our numerical simulation,
we use the following values from the experiment\cite{Anker2005PRL},
$\lambda=2\pi/k_L=783$nm, $V_0=10E_r$, and 
$\omega_\bot=2\pi \times 230$Hz.

To simulate the experiment, we prepare our initial wave function
$\psi$ to be the ground state in the combined potential
of $V\cos(x)+{1 \over 2}\omega x^2$. This is achieved by 
integrating Eq.(\ref{eq:gpe1}) with imaginary time.
In the experiment, the waveguide potential also has
a longitudinal trapping frequency at $\omega_{\parallel}=2\pi$Hz,
which is very weak and can be ignored. Therefore,
after obtaining the initial wave function, we completely remove
the longitudinal trapping and let the
wave function evolve according to the following equation
\begin{eqnarray}
\label{eq:nls_cos}
i{\partial \psi\over \partial t}=-{1 \over 2}\nabla^2\psi+
V\cos(x)\psi+g|\psi|^2\psi\,.
\end{eqnarray}
The evolutions are subsequently recorded and analyzed.

\section{Wave packet evolution}
With the above method, we have computed the evolution of the wave 
packets for different numbers of atoms in the BEC.
As indicated in Eq.(\ref{eq:nonlinear}), the number of atoms in
the BEC translates into the nonlinear parameter: larger the
atom number $N$ stronger the nonlinearity (or the repulsive
interaction). Fig.\ref{fig:width2} illustrates how the width of a 
wave packet evolves for different atom numbers. 
\begin{figure}[!b]
\includegraphics[width=\columnwidth]{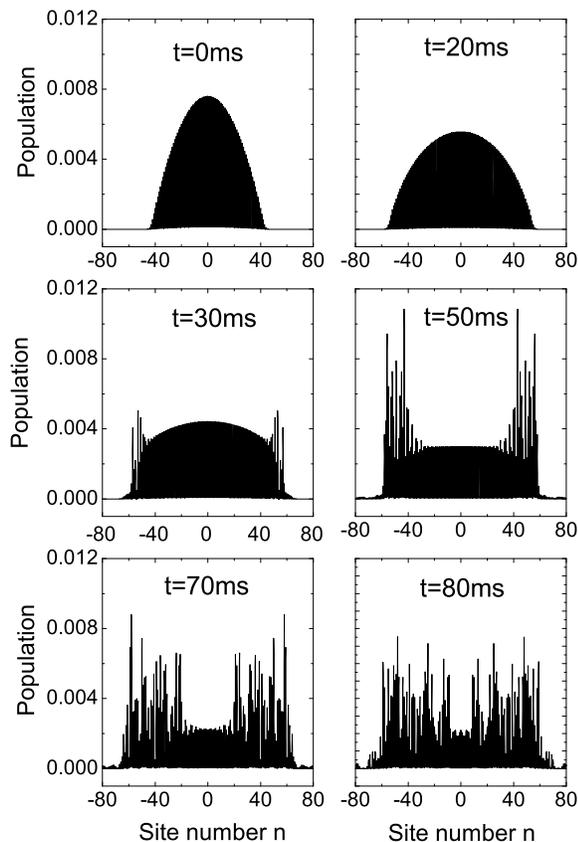}
\vspace{-0.2cm}
\caption{Time evolution of the wave packet density for
$N=5000$.}
\label{fig:t5r}
\end{figure}
It is clear from this figure that, when the BEC is dilute 
and has small atom numbers, $N\lesssim 2000$, the wave packet 
expands continuously without stopping 
as one may have expected. Also as expected, in this range, when 
the number of atoms increases, the expansion becomes faster. The 
evolution becomes very different 
when the BEC is denser. We see in Fig.\ref{fig:width2} that
for $N=3000$, the wave packet expansion slows down around 70ms
and becomes slower than the wave packet for $N=2000$ around 85ms. 
This means that we have slower expansion for a denser cloud of
repulsive interaction, a rather counter-intuitive result. As 
the cloud gets denser with more atoms, the expansion slows down
further. Around $N=5000$, there even appears a plateau where
the cloud stops expanding and becomes self-trapped as observed 
in the experiment\cite{Anker2005PRL}. 
Fig.\ref{fig:width2} illustrates a key point in this
intriguing self-trapping phenomenon: it does not
happen in a sudden and it is a gradual process. 
Before it happens, the wave packet expansion 
already slows down for higher enough densities.

What is more interesting is that, in our numerical simulation, the wave 
packet continues to expand after pausing for 30-40ms. For $N=5000$,
the expansion re-starts at $\sim 85$ms, just beyond the longest
observation time in Ref.\cite{Anker2005PRL}. Therefore, this continued 
expansion awaits for verification in future experiments.
Nevertheless, the counter-intuitive phenomenon, denser clouds
expand slower,  persists even after the expansion re-starts
as we can see in Fig.\ref{fig:width2}. In the next section,
we shall offer an explanation of this self-trapping phenomenon
and explain why it is only temporary. Just out of curiosity,
we also computed the case of very large atom number $N=50,000$;
we find that the wave packet is almost never
seen to spread. This is likely due to that the self-trapping 
lasts too long to be observed in our numerical simulations.
\begin{figure}[!b]
\includegraphics[width=\columnwidth]{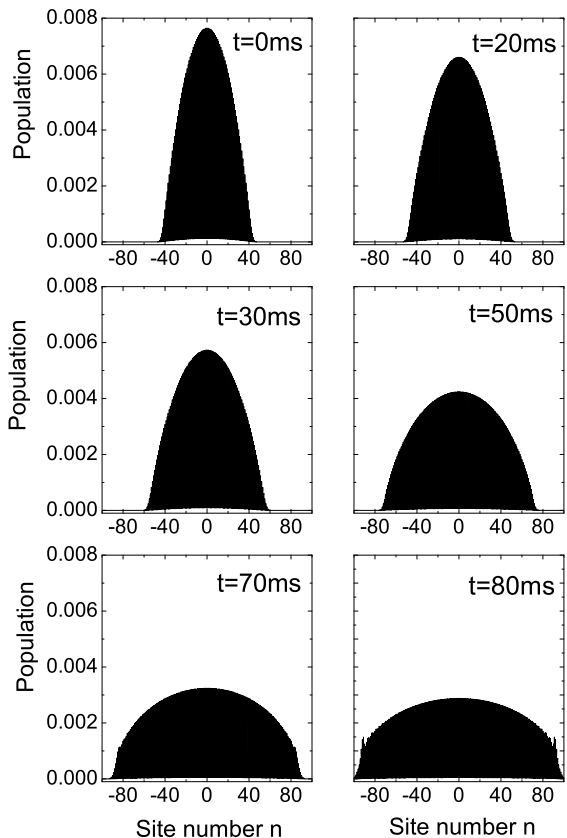}
\caption{Time evolution of the wave packet density for
$N=2000$ before $t=80$ms.}
\label{fig:t2r}
\end{figure}

\begin{figure}[!htb]
\includegraphics[width=\columnwidth]{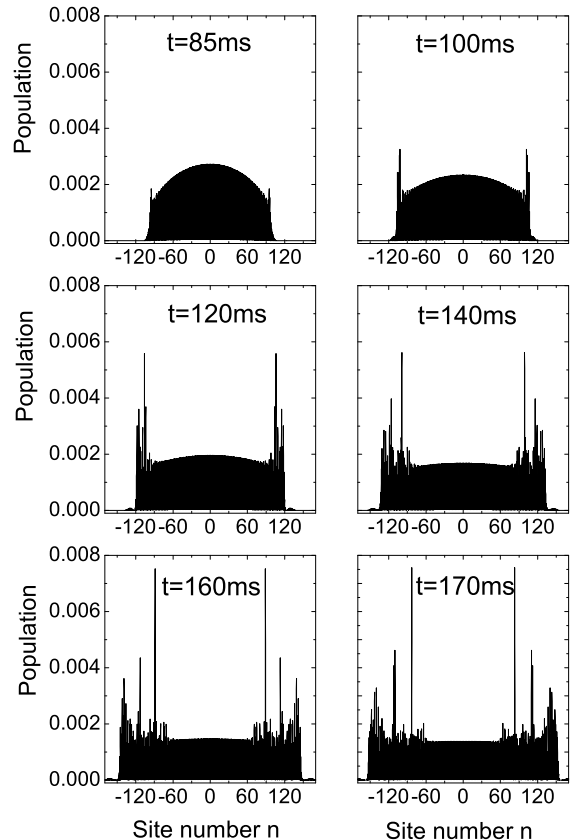}
\caption{Time evolution of the wave packet density for
$N=2000$ after $t=80$ms.}
\label{fig:t2r2}
\end{figure}

Shown in Fig.\ref{fig:t5r} are some snapshots of the time evolution 
of the wave packet for $N=5000$. Around $t=30$ms, steep edges are 
seen growing pronounced on both sides of the wave packet. 
Moreover, in the subsequent evolution, the positions of the 
steep edges do not move out any further. When compared with 
Fig.\ref{fig:width}, it is clear that this appearance of the 
steep edges coincides with the non-spreading of the wave packet.
Therefore, it seems that the steep edges seen in 
Fig.\ref{fig:t5r} signal the emergence of the self-trapping 
as suggested in Ref.\cite{Anker2005PRL}. 
The following results indicate otherwise.

Fig.\ref{fig:t2r} shows the time evolution of the
wave packet for $N=2000$. Before 80ms (about the longest
experimental observation time in Ref.\cite{Anker2005PRL}), 
there are no steep edges. However, around 85ms, the steep edges
begin to appear and grow more and more pronounced
as the evolution goes on. What is different from
$N=5000$ is that these steep edges continue to move
out during the time evolution and the width of the wave packet
also grows with time. There is no self-trapping.
This clearly shows that the steep edges do not necessarily
lead to the self-trapping of the wave packet.

\begin{figure}[!tb]
\includegraphics[width=\columnwidth]{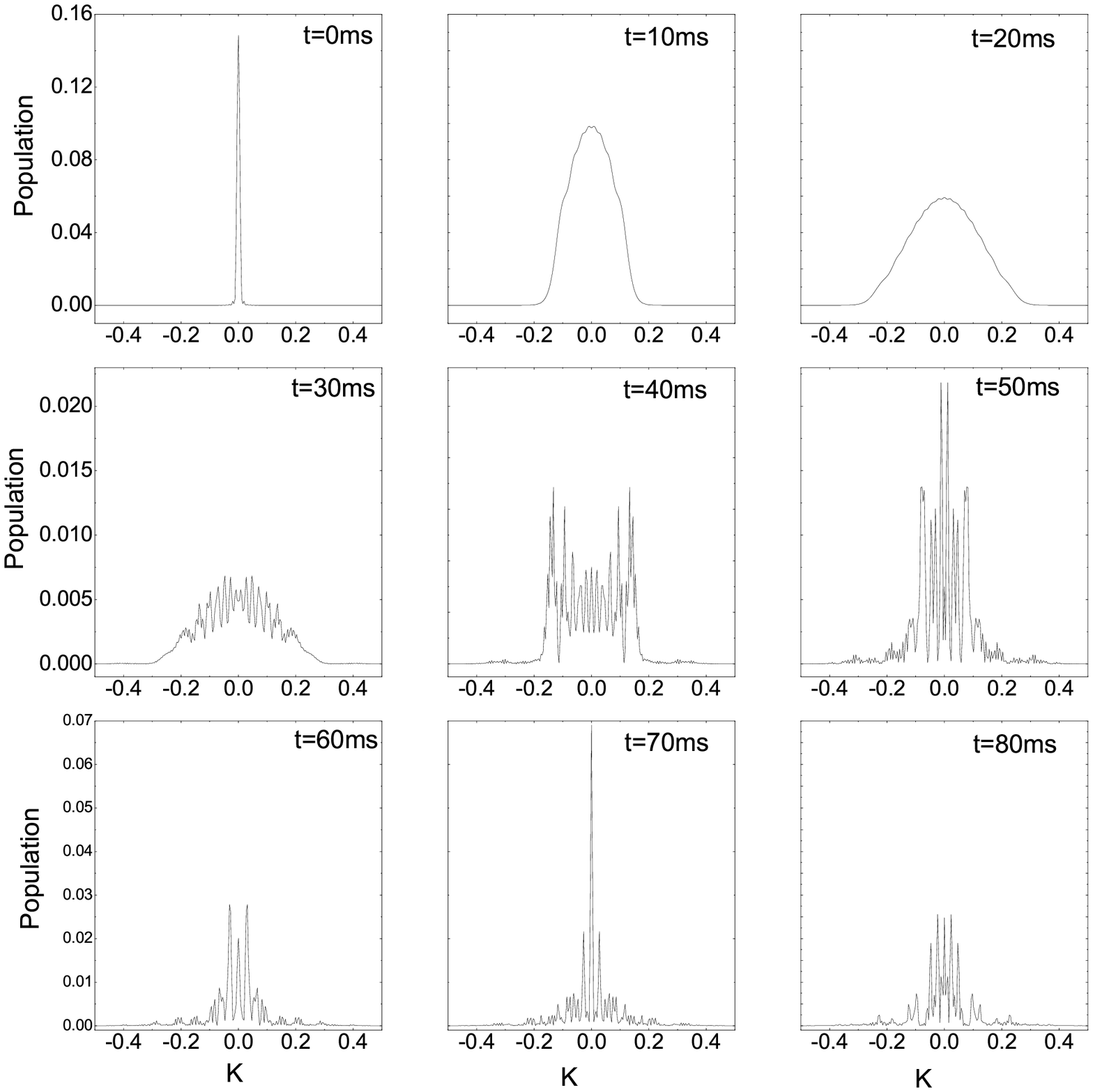} \vspace{-0.2cm}
\caption{Time evolution of the wave packet  in the quasi-momentum 
space for $N=5000$.}
\label{fig:t5k}
\end{figure}

We have also computed how the wave packet evolves in
the quasi-momentum space. To achieve this,
we expand the wave packet in terms of 
the Bloch waves belonging to the lowest Bloch band
of the linear system with the periodic potential
$\cos x$. The results are plotted in Fig.\ref{fig:t5k},
where we do not see a large population around $k=1/4$.
Therefore, the link between the formation of steep edges
and the population at $k=1/4$ believed in Ref.\cite{Anker2005PRL}
is not established here. 

\begin{figure}[!hb]
\includegraphics[width=7cm]{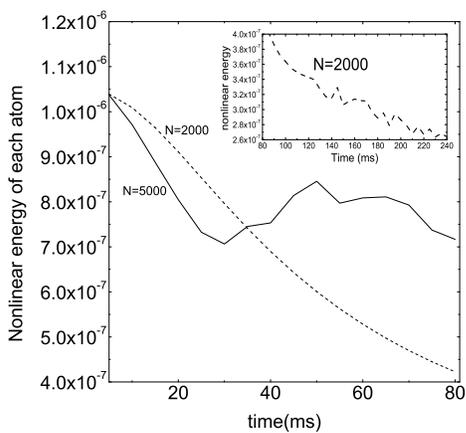} \vspace{-0.2cm}
\caption{The dynamic of nonlinear energy of each atom for $N=2000$
(dashed)and $N=5000$ (solid).}
\label{fig:neng}
\end{figure}

Another quantity that can be used to characterize the 
self-trapping phenomenon is the nonlinear energy of
the BEC, which is given by $\int \frac{g}{2}|\psi(x,t)|^4 dx$.
If the wave packet expands continuously, the nonlinear energy
should decrease with the expansion. If there is self-trapping,
i.e., the wave packet stops to grow, then the nonlinear energy
should remain largely constant. Indeed, this is the
case as shown in Fig.\ref{fig:neng}, where the nonlinear
energy does not decrease and only fluctuates slightly when 
the self-trapping occurs.

\section{relationship to   the self-trapping in a double-well}
It has been known for a while that the self-trapping also occurs 
for a BEC in a double-well potential
\cite{Milburn1997PRA,Smerzi1997PRL,WangGF2005,Albiez2005PRL}.
It is then natural to ask whether the self-trappings
in these two different systems are related to
each other. Our analysis shows that these two are closely
related. 
\begin{figure}[!b]
\includegraphics[width=\columnwidth]{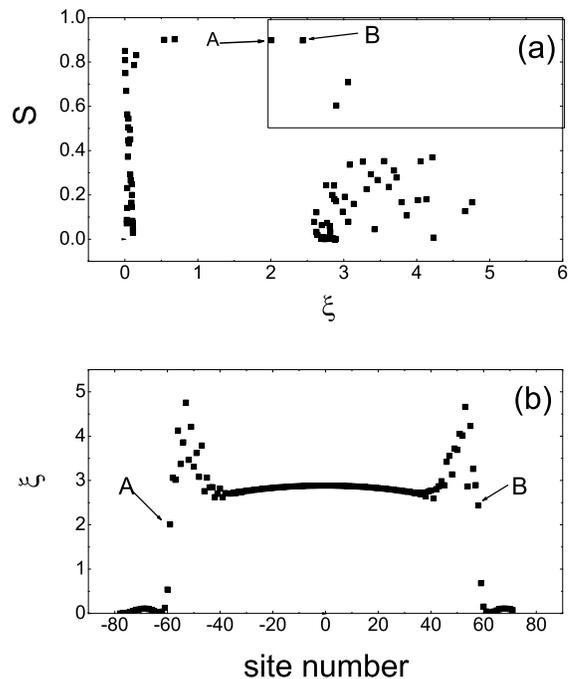} \vspace{-0.2cm}
\caption{The values of $\xi$ and $s$ of a wave packet 
at $t=40$ms for $N=5000$. (a) The values of $\xi$
at different wells (The $\xi$ at the $n$th well is
for the double wells composed of the $n$th and $n+1$th
wells. (b) Phase diagram (or distribution) of $\xi$
and $s$ for this wave packet. The self-trapping occurs
at $t=40$ms. }
\label{fig:xis5k}
\end{figure}
This was already noticed in Ref.\cite{Anker2005PRL} based
on numerical results with a tight-binding approximation of the GP equation;
here we offer a more detailed analysis with numerical results
with the full GP equation. Our analysis leads to a possible explanation
why the self-trapping observed in the optical lattice is only temporary
and how the steep edges form. 
For the sake of self-containment and introducing new parameters,
we briefly review the self-trapping in the double-well
situation.

\begin{figure}[!b]
\includegraphics[width=\columnwidth]{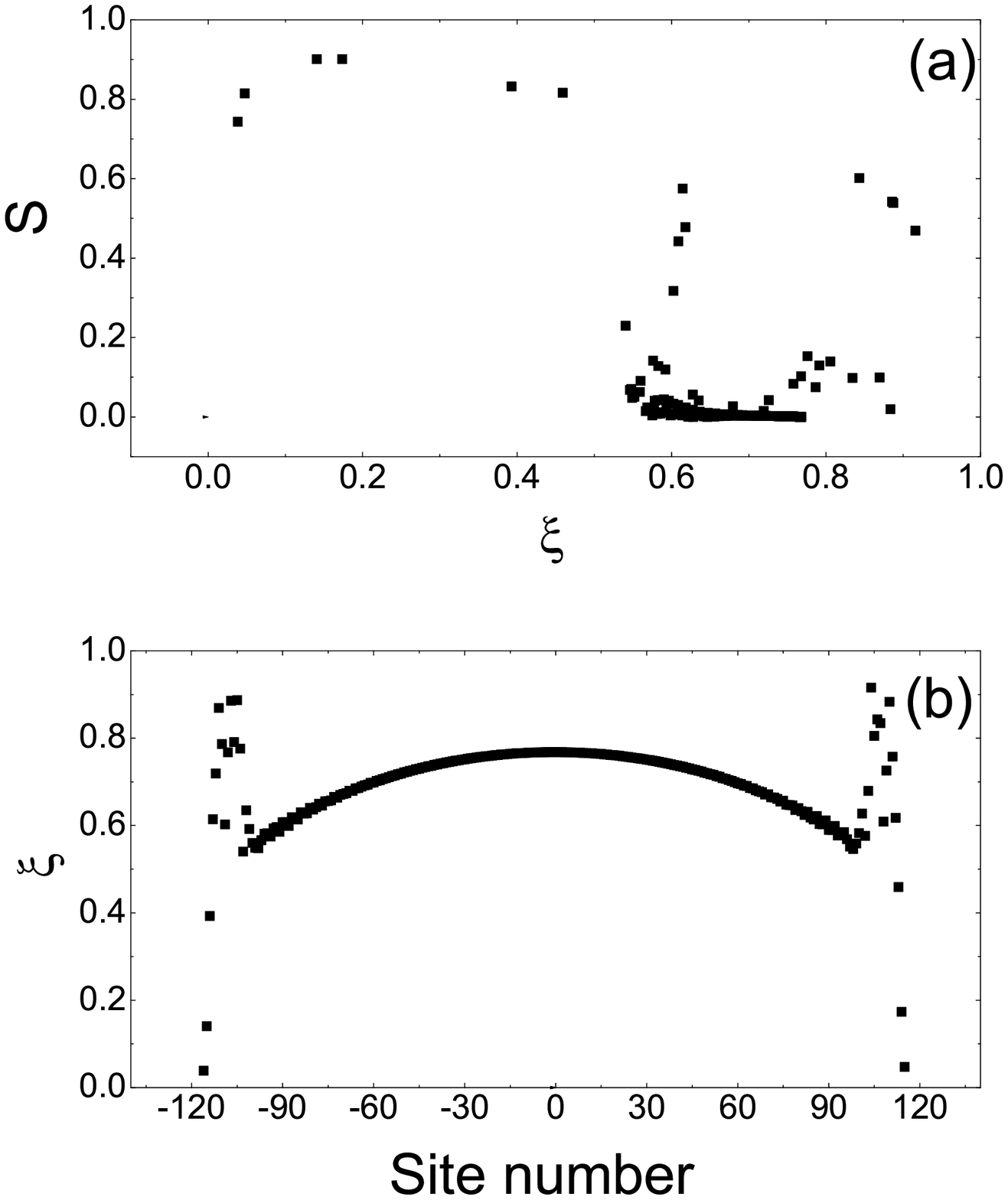} \vspace{-0.2cm}
\caption{The values of $\xi$ and $s$ of a wave packet 
at $t=110$ms for $N=2000$. (a) The values of $\xi$
at different wells (The $\xi$ at the $n$th well is
for the double wells composed of the $n$th and $n+1$th
wells. (b) Phase diagram (or distribution) of $\xi$
and $s$ for this wave packet. The time 110ms is
when the steep edges start to appear. }
\label{fig:xis2k}
\end{figure}

The Hamiltonian governing the dynamics of a BEC in a double-well
potential can be written as\cite{Milburn1997PRA,Smerzi1997PRL,WangGF2005},
\begin{equation}
\label{eq:ham}
H_{classical}= - {c\over 2}s^2+v \sqrt{1-s^2}\cos\theta,
\end{equation}
where $\theta=\theta_b-\theta_a$ is the relative phase
between the two wells $a$ and $b$ while $s$ is the fractional 
population difference $s=(N_b-N_a)/(N_a+N_b)$ with 
$N_a$ and $N_b$ being the number of atoms in wells $a$ 
and $b$, respectively. 
Previous studies \cite{Milburn1997PRA,Smerzi1997PRL,WangGF2005}
show that there are two types of self-trapping in this system, 
depending on the ratio $\xi= c/v$ and the
population difference $s$. They are: (1) If $1<\xi<2$ and 
the relative phase $\theta$ is around $\pi$,
then the self-trapping occurs when $s>0.5$.
This is called 'oscillation type' self-trapping.
(2) If $\xi>2$ and $s>0.5$, another type of self-trapping
emerges with the relative phase $\theta$ between the
two wells increasing with time. Therefore, it is called 
'running phase type' self-trapping.

Any pair of neighboring wells in the optical lattice can be 
viewed as a double-well. To establish the link between
the self-trappings in the optical lattice and the double-well,
we need to compute $\xi$ and $s$ for each pair of the neighboring 
wells in the optical lattice for a given wave packet.
The details of how $\xi$ is 
computed for neighboring wells in an optical lattice can be 
found in Appendix A.

\begin{figure}[!tb]
\includegraphics[width=\columnwidth]{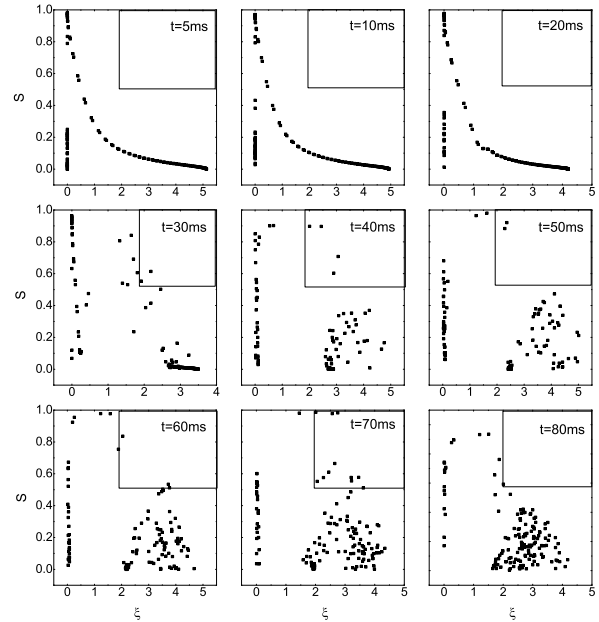} \vspace{-0.2cm}
\caption{Phase diagrams (or distributions) of $s$ and $\xi$ for 
wave packets at different evolution times. $N=5000$.}
\label{fig:phase}
\end{figure}

Fig.\ref{fig:xis5k} shows one set of such calculations for 
a wave packet with $N=5000$ at $t=40$ms, which is the time
when the self-trapping happens. It is clear from 
Fig.\ref{fig:xis5k}(a) that there are four pairs of double-wells
whose $\xi$ and $s$ satisfy the condition for the ``running 
phase type'' self-trapping in the double-well system. 
Furthermore, two of these four pairs, marked by A and B 
in Fig.\ref{fig:xis5k}, are located right at the two edges
of the wave packet. (The other two are just nearby,
for clarity we do not mark them.) It seems to suggest that these two 
self-trapped pairs of double-wells serve as two dams stopping
the flow of atoms to the outside. Therefore, the
self-trapping in the optical lattice appears just alternative
manifestation of the self-trapping in the double-well system. 

To firmly establish such a link, we have also examined the case $N=2000$, 
where there is no self-trapping. Fig.\ref{fig:xis2k} shows the values $s$ 
and $\xi$ for the wave packet with $N=2000$ at $t=110$ms. This is the time 
when the steep edges have already developed.  We see from the figure
that all the values of $\xi$ are smaller than one: the self-trapping 
conditions of the double-well system are not satisfied by any pair of
neighboring wells in the optical lattice.

The above analysis leads to the following conclusion:
when there is self-trapping in the optical lattice,
there are neighboring wells that satisfy the self-trapping
condition of the double-well system; when there is no
self-trapping in the optical lattice, any pair of 
the neighboring wells in the lattice does not satisfy
the double-well self-trapping condition. So established is
a solid link between these two self-trapping
phenomena. One intuitive way of understanding of this link is
such. Once the self-trapping happens in 
some pairs of neighboring wells around the edges
of a wave packet, these self-trapped double-wells,
behaving like ``dams'', stop the tunneling of atoms towards 
outside, causing the non-spreading of the wave packet.
In Fig.\ref{fig:phase} we have plotted a series of ``phase diagrams'',
where the distribution of the $s$-$\xi$ pairs is shown.
The square region bounded by the dashed
line in each panel is the area where the self-trapping conditions
are satisfied. We can see clearly from this figure  that the
self-trapping happens from about $t=40$ms to 80ms for
$N=5000$. We have also checked the cases of $N=7000$ 
and $N=50,000$ and reached the same conclusion.

This link not only explains why the self-trapping occurs in the
optical lattice but also offers a possible mechanism why
the self-trapping is temporarily lived. At the outmost wells,
the density of the BEC is very low and the self-trapping conditions
of the double-well system can never be satisfied. As a result,
the atoms will tunnel towards outside. The amount of atoms
tunneling out is very small and has not much effect on the evolution
of the whole cloud. However, for long evolution times, this small
amount of ``dripping'' can lead to the significant decreasing of 
atom numbers in the wells, thus destroy the self-trapping. This 
is similar to that small cracks can cause the collapse of a 
dam in a long time. 

Based on the link between these two self-trappings, 
it is also possible to understand why the BEC cloud expands slower 
around $N=3000$ than $N=2000$ seen in Fig.\ref{fig:width2}.
As one can imagine, when the cloud density increases, some pairs
of the wells will get close to satisfy these self-trapping conditions 
and eventually satisfy them. For the medium densities,e.g., $N=3000$ 
there should be a few pairs of the wells that satisfy the
conditions just barely. As a result, the self-trapping conditions can be 
easily or quickly 
destroyed by the ``dripping'' effect mentioned above. However,
as the cloud expands, the self-trapping conditions can again be
satisfied by some pairs of well further inside and then destroyed again.
This on-and-off process can dramatically lead to slowing down of
the cloud expansion. What is a pity is that this straightforward picture
is hard to be corrobarated by our numerical computation 
because the values $s$ and $\xi$ for neighboring wells can only
be computed approximately. Alternative methods may be needed to
verify this picture.

\begin{figure}[!htb]
\includegraphics[width=\columnwidth] {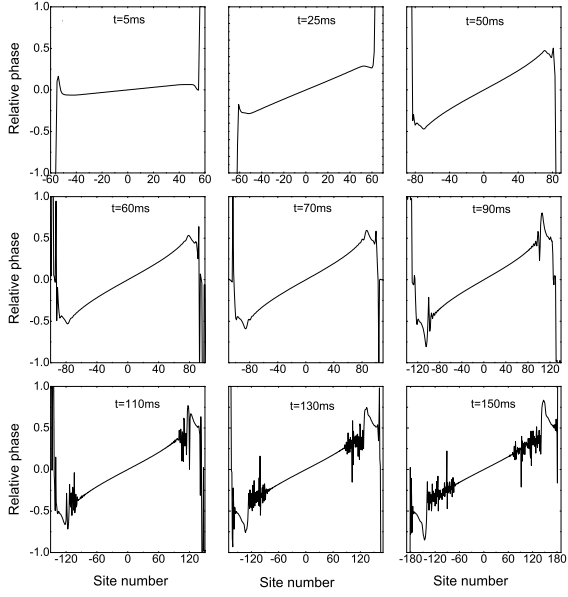} \vspace{-0.2cm}
\caption{The evolution of relative phases for each pair of neighboring wells 
in the optical lattice. $N=2000$.}
\label{fig:phase2}
\end{figure}

\section{Steep edges }
\begin{figure}[!htb]
\includegraphics[width=\columnwidth] {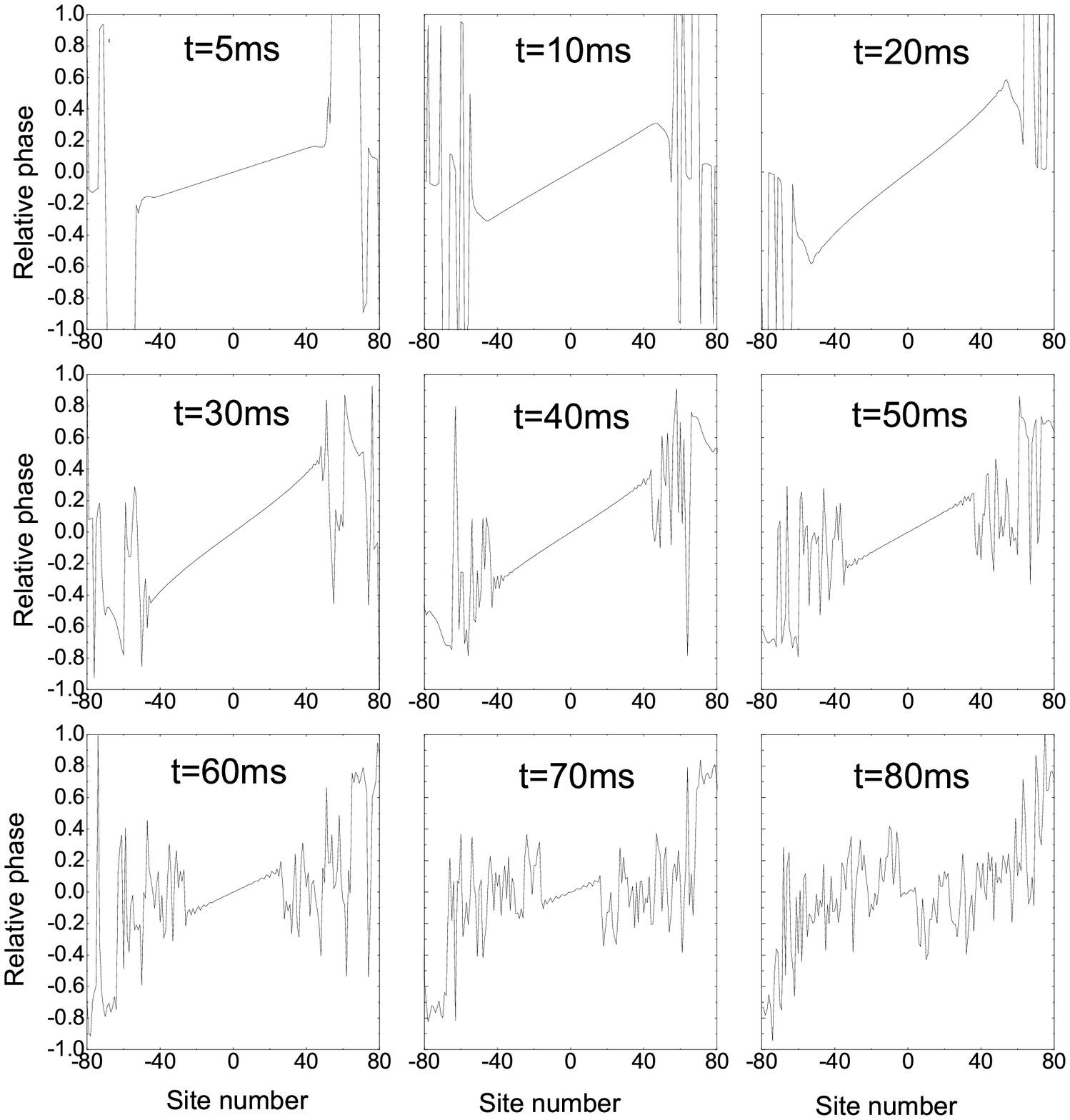} \vspace{-0.2cm}
\caption{The evolution of relative phases for each pair of neighboring wells 
in the optical lattice. $N=5000$.}
\label{fig:phase5}
\end{figure}

\begin{figure}[!htb]
\includegraphics[width=\columnwidth] {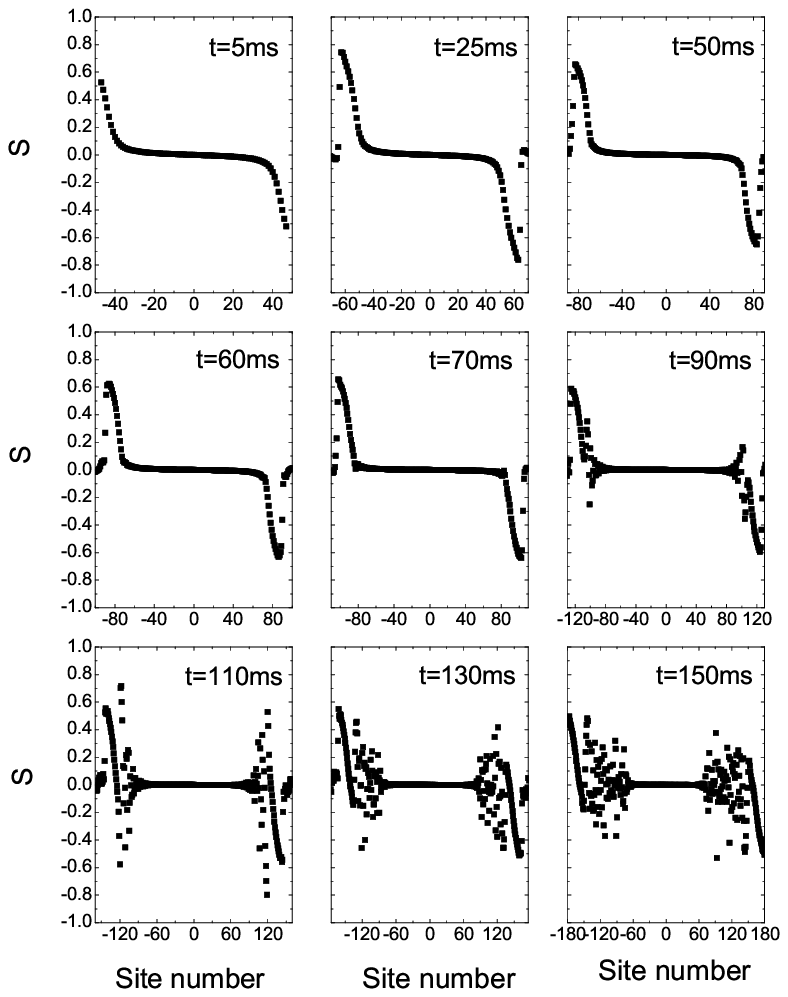} \vspace{-0.2cm}
\caption{The evolution of relative populations $s$
for each pair of neighboring wells 
in the optical lattice. $N=2000$.}
\label{fig:pop2}
\end{figure}

\begin{figure}[!htb]
\includegraphics[width=\columnwidth] {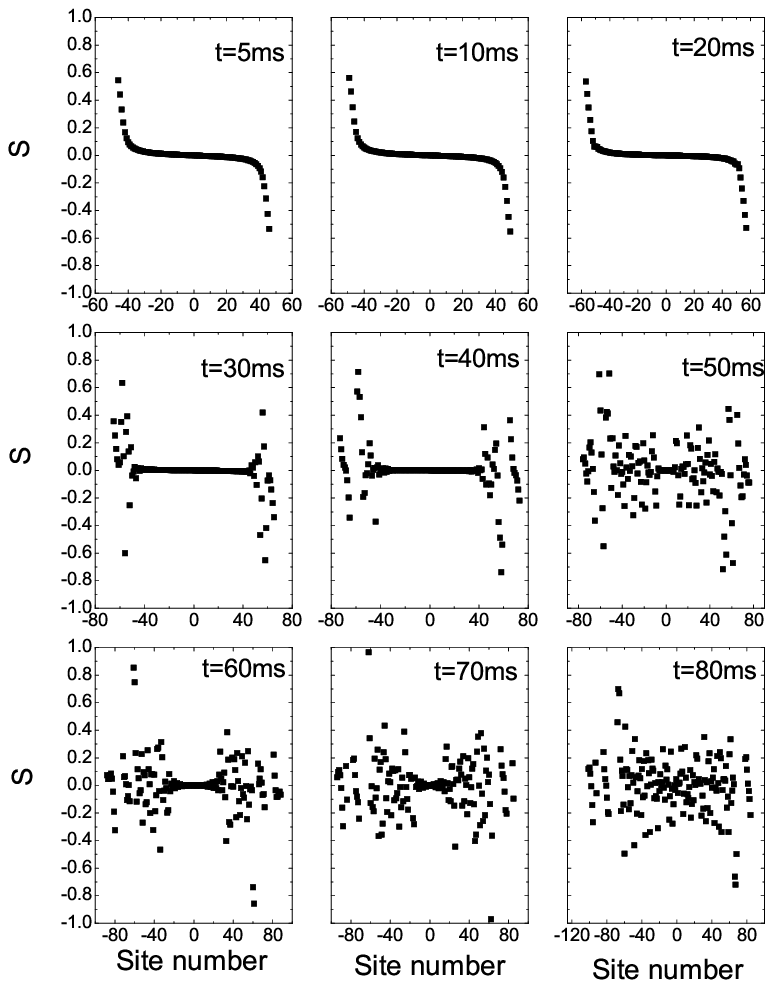} \vspace{-0.2cm}
\caption{The evolution of relative populations $s$
for each pair of neighboring wells 
in the optical lattice. $N=5000$.}
\label{fig:pop5}
\end{figure}
There is a mystery in the wave packet evolution yet to
be explained, that is, the appearance of steep edges.
Our following analysis shows that the formation of steep edges
can also be understood in terms of the BEC dynamics in
the double-well system. For future convenience, 
we write down the dynamics in the double-well system
\begin{eqnarray}
\label{eq:s}
\dot{s}&=&v \sqrt{1-s^2}\sin\theta, \\
\label{eq:theta}
\dot{\theta}&=&-cs-{v s \cos\theta \over \sqrt{1-s^2}},
\end{eqnarray}
which are derived from the Hamiltonian in Eq.(\ref{eq:ham}).

In Fig.\ref{fig:phase2}, we have plotted
how the relative phases $\theta$ evolve with time for $N=2000$. 
In our calculation, the relative 
phase $\theta$ is defined as the phase difference between
the middle points in two neighboring wells. 
Initially, the relative phase is zero for every pair of double-wells 
in the optical lattice
as indicated by a horizontal line. As the evolution goes on, the line 
of relative phase begins to incline with an increasing slope.
The tendency stops when the two end points of the line reach
$\pm \pi/2$, respectively.  This is around $t=80$ms, right when the
steep edges appear. As shown in Fig.\ref{fig:phase5}, the situation is
similar for $N=5000$. We also observe from Figs.\ref{fig:pop2}\&\ref{fig:pop5} 
that the relative populations in the middle of the wave packet
remain largely zero before the appearance of steep edges.

According to Eq.(\ref{eq:s}), at $\theta\sim \pi/2$ and $s\sim 0$,
the tunneling or transfer of atoms between these neighboring wells
is the largest. This means that there are more atoms flowing into
some particular wells than going out, thus generating steep edges.
We notice that right after the appearance of the steep edges. 
both $s$ and $\theta$ become rather ``random''. 
This is likely due to the complicated dynamics caused
by steep edges.

\section{conclusion}
With the GP equation, we have studied the wave packet 
dynamic of a BEC in a one-dimensional optical lattices.
We find an intriguing self-trapping phenomenon in
the expansion of the wave packet, agreeing with
a recent experiment\cite{Anker2005PRL}. Moreover,
we find that  the self-trapping is only temporary
and the wave packet continues to grow at long evolution
times that are beyond the current experiment\cite{Anker2005PRL}.
The analysis of our numerical results shows
that the self-trapping in the optical lattice
is closely related to the self-trapping found in the system
of a BEC in a double-well potential. We also showed
that the steep edges appearing the wave packet evolution
do not necessarily lead to self-trapping and they can
also be understood in terms of the dynamics in the
double-well systems.

\begin{acknowledgments}
B. Wang is supported by the National
Natural Science Foundation of China under Grant No. 60478031.
J. Liu is supported by the NSF of China (10474008), the 973 project
(2005CB3724503), and the 863 project (2004AA1Z1220). 
B. Wu is supported by the ``BaiRen'' program of the 
Chinese Academy of Sciences and the 973 project (2005CB724500).
\end{acknowledgments}

\appendix
\section {Computation of $\xi$ in optical lattices}
We choose one well and its neighbor as a double well trap. The GP
equation is as Eq. (\ref{eq:nls_cos}) except the potential is replaced
by $V(x)=V\cos(x)$ for $|x|<2\pi$ and $V(x)=V$ for
$|x|>2\pi$. Then we write the wave function as a two-mode wave
function: $\phi = au_1(x) + bu_2(x)$ and let $|a|^2=N_a/(N_a+N_b)$
and $|b|^2=N_b/(N_a+N_b)$ ($N_a$ is the particle number in well a
and $N_b$ the particle number in well b). Plugging the double mode
wave function into the GP equation with potential $V(x)$ and
using the tight-binding approximation, we obtain the effective
Hamiltonian:
\begin{equation}
H_{eff}=-c/2(|a|^2-|b|^2)^2+v (a^*b+ab^*),
\end{equation}
where the parameter $c=(N_a+N_b)Ng\int|u_1(x)|^4dx$ and
$v=N\int(-{1 \over2}\triangledown u_1^*(x)\triangledown
u_2(x)+ V'(x)u_1^*(x)u_2(x))dx$. The wave function $u_1(x)$ (or
$u_2(x)$) is obtained as a ground state from Eq.(\ref{eq:nls_cos}) with single well:
$\tilde{V}(x)=V\cos(x)$ for $0<x<2\pi$ and $\tilde{V}(x)=V$ 
for other $x$ values.


\end{document}